\documentstyle[12pt,amssymb]{article}
\catcode`\@=11

\catcode`\@=11

\def\section{\@startsection{section}{1}{\z@}{3.5ex plus 1ex minus
   .2ex}{2.3ex plus .2ex}{\large\bf}}

\def\ps@headings{\def\@oddfoot{}\def\@evenfoot{}
\def\@oddhead{\hbox{}\hfill
        \makebox[.5\textwidth]{\raggedright\ignorespaces --\thepage{}--
        \hfill }}
\def\@evenhead{\@oddhead}
\def\subsectionmark##1{\markboth{##1}{}}
}

\ps@headings

\catcode`\@=12
\catcode`\@=12

\relax

\def\figcap{\section*{Figure Captions\markboth
        {FIGURECAPTIONS}{FIGURECAPTIONS}}\list
        {Fig. \arabic{enumi}:\hfill}{\settowidth\labelwidth{Fig. 999:}
        \leftmargin\labelwidth
        \advance\leftmargin\labelsep\usecounter{enumi}}}
 \relax
\def\tablecap{\section*{Table Captions\markboth
        {TABLECAPTIONS}{TABLECAPTIONS}}\list
        {Table \arabic{enumi}:\hfill}{\settowidth\labelwidth{Table 999:}
        \leftmargin\labelwidth
        \advance\leftmargin\labelsep\usecounter{enumi}}}
 \relax
\def\reflist{\section*{References\markboth
        {REFLIST}{REFLIST}}\list
        {[\arabic{enumi}]\hfill}{\settowidth\labelwidth{[999]}
        \leftmargin\labelwidth
        \advance\leftmargin\labelsep\usecounter{enumi}}}
 \relax

\catcode`\@=11

\def\marginnote#1{}
\newcount\hour
\newcount\minute
\newtoks\amorpm
\hour=\time\divide\hour by60
\minute=\time{\multiply\hour by60 \global\advance\minute by-
\hour}
\edef\standardtime{{\ifnum\hour<12 \global\amorpm={am}%
    \else\global\amorpm={pm}\advance\hour by-12 \fi
    \ifnum\hour=0 \hour=12 \fi
    \number\hour:\ifnum\minute<100\fi\number\minute\the\amorpm}}
\edef\militarytime{\number\hour:\ifnum\minute<100\fi\number\minute}
\def\draftlabel#1{{\@bsphack\if@filesw {\let\thepage\relax
  \xdef\@gtempa{\write\@auxout{\string
    \newlabel{#1}{{\@currentlabel}{\thepage}}}}}\@gtempa
    \if@nobreak \ifvmode\nobreak\fi\fi\fi\@esphack}
     \gdef\@eqnlabel{#1}}
\def\@eqnlabel{}
\def\@vacuum{}
\def\draftmarginnote#1{\marginpar{\raggedright\scriptsize\tt#1}}
\def\draft{\oddsidemargin -.5truein
        \def\@oddfoot{\sl preliminary draft \hfil
        \rm\thepage\hfil\sl\today\quad\militarytime}
        \let\@evenfoot\@oddfoot \overfullrule 3pt
        \let\label=\draftlabel
        \let\marginnote=\draftmarginnote
\def\@eqnnum{(\theequation)\rlap{\kern\marginparsep\tt\@eqnlabel}%
\global\let\@eqnlabel\@vacuum}  }
\def\preprint{\twocolumn\sloppy\flushbottom\parindent 1em
        \leftmargini 2em\leftmarginv .5em\leftmarginvi .5em
        \oddsidemargin -.5in    \evensidemargin -.5in
        \columnsep 15mm \footheight 0pt
        \textwidth 250mmin      \topmargin  -.4in
        \headheight 12pt \topskip .4in
        \textheight 175mm
        \footskip 0pt

\def\@oddhead{\thepage\hfil\addtocounter{page}{1}\thepage}
        \let\@evenhead\@oddhead \def\@oddfoot{} \def\@evenfoot{}
}
\def\titlepage{\@restonecolfalse\if@twocolumn\@restonecoltrue\onecolumn
     \else \newpage \fi \thispagestyle{empty}\c@page\z@
        \def\thefootnote{\fnsymbol{footnote}} }
\def\endtitlepage{\if@restonecol\twocolumn \else  \fi
        \def\thefootnote{\arabic{footnote}}
        \setcounter{footnote}{0}}  
\catcode`@=12
\relax

\def\ps@headings{\def\@oddfoot{}\def\@evenfoot{}
\def\@oddhead{\hbox{}\hfill
        \makebox[.5\textwidth]{\raggedright\ignorespaces --\thepage{}--
        \hfill }}
\def\@evenhead{\@oddhead}
\def\subsectionmark##1{\markboth{##1}{}}
}

\ps@headings

\relax

\def\firstpage#1#2#3#4#5#6{
\begin{document}
\newcommand{\newc}{\newcommand}
\newc{\ra}{\rightarrow}
\newc{\lra}{\leftrightarrow}
\newc{\beq}{\begin{equation}}
\newc{\eeq}{\end{equation}}
\newc{\bea}{\begin{eqnarray}}
\newc{\eea}{\end{eqnarray}}
\begin{titlepage}
\nopagebreak
\title{\begin{flushright}
 \vspace*{-0.8in}
{ \normalsize 
 hep-th/9908106}\\
{\normalsize CERN-TH-99-251}\\
\end{flushright}
\vfill
{#3}}
\author{\large #4 \\[1.0cm] #5}
\maketitle
\vskip -7mm
\nopagebreak
\begin{abstract}
{\noindent #6}
\end{abstract}
\vfill
\begin{flushleft}
\vspace*{.6cm} {\normalsize
August 1999}\\[-9mm]
\end{flushleft}
\thispagestyle{empty}
\end{titlepage}}

\def\simlt{\stackrel{<}{{}_\sim}}
\def\simgt{\stackrel{>}{{}_\sim}}
\date{}
\firstpage{3118}{IC/95/34} {\large\bf The Matrix model and the non-commutative 
geometry\\ of the supermembrane}
{E.G. Floratos$^{\,a,b,*}$ and G.K. Leontaris$^{\,c,d,**}$}
{\normalsize\sl
$^a$Institute of Nuclear Physics,  NRCS Demokritos,
{} Athens, Greece\\
\normalsize\sl
$^b$ Physics Department, University of Iraklion,
Crete, Greece\\
\normalsize\sl
$^c$Theoretical Physics Division, Ioannina University,
GR-45110 Ioannina, Greece\\
\normalsize\sl
$^d$ CERN, Theory Division, 1211 Geneva 23, Switzerland
}
{This is a short note on the relation of the Matrix
model with the non-commutative geometry of the 11-dimensional
supermembrane. We put forward the idea that M-theory is 
described by the 't Hooft topological expansion of the Matrix
model in the large $N-$limit where all topologies of membranes 
appear. This expansion can faithfully be represented by the Moyal 
Yang-Mills theory of membranes. We discuss this conjecture
in the case of finite $N$, where the non-commutative geometry
of the membrane is given be the finite quantum mechanics. 
The use of the finite dimensional representations
of the Heisenberg group reveals the cellular structure of a
toroidal supermembrane on which the Matrix model appears as
a non--commutatutive Yang--Mills theory. The Moyal star product
on the space of functions in the case of rational values of
the Planck constant $\hbar$ represents exactly this cellular structure.
We also discuss the integrability of the instanton sector
as well as the topological  charge and the corresponding 
Bogomol'nyi bound.
}
\vspace*{2cm}
$^*$ manolis@timaios.nuclear.demokritos.gr\\
$^{**}$leonta@mail.cern.ch
\newpage
\section{Introduction}

One of the basic ingredients of M--threory\cite{Witten:1996im,Townsend:1996af} 
is the eleven dimensional (11-d) supermembrane for which some years
ago\cite{Bergshoeff:1987cm} a consistent action has been written in a general
background of 11-d supergravity. The supermembrane has a
uniquely defined self--interaction, which comes in contrast to the
superstring, from a infinite dimensional gauge symmetry
apparent in the light-cone gauge as the area-preserving
diffeomorphisms  on the surface of the membrane.

Because  of the absence of the dilaton field for the
supermembrane, there is no topological expansion over all possible three
manifolds analogous to the string case.
 The supermembrane, due to its unique
self-interaction, is possible to break into other supermembranes
so in a sense is already a second quantized theory
 but up to now there is no consistent perturbative
expansion. In
the light-cone gauge, and flat space-time, there are two classes
of membrane vacua, points and tensionless strings, so a low--energy
effective field theory of supermembrane massless excitations would
be either eleven-dimensional supergravity or a field theory for
tensionless strings. Hopefully,
recent efforts for understanding the coupling
of 11-d supergravity with the supermembrane will help 
to the construction of its effective low energy field
theory\cite{deWit:1999rh}.

In this letter, we present arguments that the Matrix model
\cite{Banks:1997vh,Ishibashi:1996xs,Polychronakos:1997fw}
describes the non-commutative geometry of the 11-d supermembrane,
and M theory is the 't Hooft topological expansion of the Matrix model.
We demostrate the existence of a topological charge and the corresponding
Bogomol'nyi bound and we discuss  the integrability of the instanton sector.

\section{Non-commutative geometry of the membrane}

It is a well known fact  that the Matrix
model\cite{Banks:1997vh,Ishibashi:1996xs,Polychronakos:1997fw}
was  one of the first ideas for the  study of the dynamics 
of the bosonic membrane in the light-cone frame and in the approximation
of finite number of oscillation  modes \cite{Goldstone,Hoppe}.
 The true dynamics would be
determined by taking the limit of infinite number of modes.
In the finite mode approximation the Hamiltonian of the
membrane  is exactly the same with $SU(N)$ Yang-Mills (YM) classical mechanics
and this system is known to possess interesting chaotic 
dynamics\cite{Matinian:1981dj}
and a discrete spectrum at the level of quantum mechanics (QM)~\cite{Simon:1983jy}.
Later on, Townsend et al\cite{Bergshoeff:1987cm,Bergshoeff:1990hw} 
discovered the supermembrane Lagrangian
in 11 dimensions and the finite mode truncation, as was expected, is 
described by the Hamiltonian of the supersymmetric $SU(N)$ YM mechanics. 
It was found that the quantum mechanical spectrum of this model is 
continuous;  at that time this was considered to be the end of the supermembrane
as a fundamendal object replacing the superstring and producing all the
low energy physics that could be useful for the unification of gauge and
gravitational forces~\cite{deWit:1988ig,deWit:1989ct}. 

In ref.~\cite{Floratos:1989au} the question of a deeper origin of the $SU(N)$ YM
classical mechanics as an approximation of the membrane dynamics was 
considered and it was found that $SU(N)$ represents the  Lie
algebra of the finite Heisenberg group, which acts on a discretized  
membrane representing a toroidal discrete phase space.
 The membrane coordinates
are approximated by $N\times N$ matrices 
(YM gauge fields), which represent collectively $N^2$ number of points
in the target space. The large $N$--limit
to reproduce the continuous surface of the membrane, should be such that all the 
positions of the $SU(N)$ matrices are filled up in a
 continuous way and this limit
has not been expressed, up to now, in a mathematically consistent way~\cite{FInew}.
The non-commutative geometry of the discrete membrane is generated by the
finite and discrete Heisenberg group and the space of functions on the 
surface of the membrane is the algebra of $N\times N$ complex matrices.

In  modern language~\cite{Connes:1998cr} 
the $SU(N)$ YM classical
mechanics is  the YM theory on  non-commutative 2-torus.
 It is interesting that the torus
compactified Matrix model is equivalent with the M--theory compactification in a
constant antisymmetric background  gauge 
field.
In this case, the Matrix model description becomes that of a gauge
theory on a  non--commutative 
torus\cite{Connes:1998cr,Hofman:1998iv,Douglas:1999ge,Douglas:1998fm,Ho:1998xh}.

It is well known that the usual Quantum Mechanics can be represented on
 functions of the phase-space variables, with the Moyal 
bracket\cite{Moyal}\footnote{For a recent discussion see 
\cite{Zachos:1999wn,Fairlie:1997vj}
and references therein.}
replacing the classical Poisson bracket.
Recently the vertex operators  of open strings  in an external antisymmetric 
gauge field $B_{\mu\nu}$ were found to obey non-commutative 
relations of the Weyl type, which  
  induces a Moyal bracket structure
on the space of functions on the string momenta
~\cite{Schomerus:1999ug}\footnote{For recent discussions see\cite{string99}.}.

\section{The Heisenberg-Weyl group and the Moyal bracket}

To start with, we introduce the irreducible representations of the finite
 Heisenberg group appropriate for the Matrix model non-commutative
geometry of a toroidal membrane. 
The Hilbert space ${\Bbb H}_{\Gamma}$ of the wave functions on the torus
$\Gamma={\Bbb C}/{\Bbb L}$ of complex modulus $\tau=\tau_1+{\imath}\tau_2$,
where ${\Bbb L}$ is the integer lattice, ${\Bbb L}=\{m_1+\tau m_2|(m_1,m_2)\in
{\Bbb Z}\times{\Bbb Z}\}$ is defined as the space of functions of complex
argument $z=x+{\imath}y$:
\begin{equation}
f(z)=\sum_{n\in{\Bbb Z}}c_n{\mathrm e}^{{\imath}\pi n^2\tau+2\pi{\imath}nz}
\end{equation}
with norm
\begin{equation}
||f||^2=\int {\mathrm e}^{-2\pi y^2/\tau_2}|f(z)|^2dxdy,\,\,\,\tau_2>0.
\end{equation}
Consider the  subspace ${\Bbb H}_{N}(\Gamma)$ of ${\Bbb H}_{\Gamma}$
with periodic Fourier coefficients $\{c_n\}_{n\in{\Bbb Z}}$ of period $N$:
\begin{equation}
\label{fourier}
c_n=c_{n+N}\,\,\,n\in {\Bbb Z},\,\,N\in{\Bbb N}.
\end{equation}
The space   ${\Bbb H}_{N}(\Gamma)$ is $N$-dimensional and there is a discrete
Heisenberg group,
with generators ${\cal S}_{1/N}$ and ${\cal T}_{1}$
acting as~\cite{cartier,bi}
\begin{eqnarray}
\label{STl}
({\cal S}_{1/N}f)(z) &=& \sum_{n\in{\Bbb Z}}c_n
{\rm e}^{2\pi{\imath}n/N}{\rm e}^{2\pi{\imath}nz+\pi{\imath}n^2\tau}\nonumber\\
 & & \nonumber\\
({\cal T}_{1}f)(z) &=& \sum_{n\in{\Bbb Z}}c_{n-1}
{\rm e}^{2\pi{\imath}nz+\pi{\imath}n^2\tau},\,\,\,c_n\in{\Bbb C}.
\end{eqnarray}
On the $N$-dimensional subspace of vectors $(c_1,\ldots,c_N)$ the two
generators are represented by $N\times N$ matrices, $Q,P$
\begin{eqnarray}
\label{gener}
({\cal S}_{1/N})_{n_1,n_2}=Q_{n_1,n_2}=\omega^{(n_1-1)}\delta_{n_1,n_2},
\;\;
({\cal T}_{1})_{n_1,n_2}=P_{n_1,n_2}=\delta_{n_1-1,n_2},
\nonumber\\
\end{eqnarray}
with $\omega=\exp(2\pi{\imath}/N)$. They satisfy the   Weyl relation 
$QP=\omega PQ$.

The Heisenberg group elements are defined as
\begin{equation}
\label{heisel}
{\cal J}_{r,s}=\omega^{r\cdot s/2}P^rQ^s.
\end{equation}
These $N\times N$ matrices are unitary 
${\cal J}_{r,s}^{\dagger} = {\cal J}_{-r,-s}$
and periodic with period $N$, i.e. 
${\cal J}_{r,s}^N = 1$.
They realize a projective representation of the discrete 
translation group $Z_N\times Z_N$:
\bea
{\cal J}_{r,s} {\cal J}_{r',s'}= \omega^{(r's-r s')/2} {\cal J}_{r+r',s+s'}
\eea

In ref\cite{Banks:1997vh}  the finite  $N$-Matrix model is considered
as a non-commutative  QM system (see also  
\cite{Floratos:1989au}),  but  the canonical commutation relations
were not represented through   the finite Heisenberg group basis
${\cal J}_{r,s}$.
 It is possible to define finite dimensional matrices
$\hat p, \hat q$  such that $Q=e^{\imath\hat{q}}$ and $P = e^{-\imath \hat{p}}$
\bea
\hat{q}_{ij}= \frac{2\pi}{N}(s+1-i)\delta_{ij},
&
\hat{p}_{ij}=-\imath\frac{\pi}N\frac{(-1)^{(i-j)}}{sin \frac{\pi}N (i-j)}
\label{mom}
\eea
where $N=2 s +1$ and $s$ is an integer. Here
we have shifted by $s$ rows and columns of $Q$ and $P$ matrices
defined in relations (\ref{gener}).   
 These matrices satisfy new Heisenberg commutation relations,
which have a very simple form\cite{Floratos:1997wg}
\beq
-\imath [\hat{q},\hat{p}]_{ij}= \frac{2\pi}N
 \frac{  \frac{\pi}N (i-j) (-1)^{(i-j)}}{
\sin{\frac{\pi}N (i-j)}}
\label{com}
\eeq
when $i\not= j$ and zero when $i=j$.  The matrix $\hat q$ satisfies
the torus compactification relations of the Matrix model, 
with corrections due to their finite size
\bea
P^{-1}\hat q P = \hat q + \frac{2 \pi}{N} I_N - 2\pi I_0,
\eea
where $I_N$ is the $N\times N$ identity matrix and
$I_0$ the $N\times N$ diagonal matrix with diagonal elements
$\{1,0,\dots , 0\}$.

The bosonic part of the matrix model is the $SU(N)$ YM classical
mechanics and the gauge fields are linear combinations of the
elements ${\cal J}_{r,s}$, i.e.,
\begin{equation}
A_{l}(t)= \sum_{r,s = 0}^{N-1}A_{l}^{r,s} J_{r,s},\;\; l=1,\dots, d-1
\label{Alt}
\end{equation}
which can be considered as coherent states of the discrete and
finite  toroidal phase-space $N\times N$ lattice.
The $A_{l}$ matrices are the non-commutative coordinates of
the discrete membrane in $d-1$ dimensions.

There is another representation of the standard 
quantum mechanics on the space of functions
of  the phase-space variables.
This is the unique deformation of the Poisson bracket, the
 Moyal bracket~\cite{Moyal,Fairlie:1997vj}
\begin{equation}
\{\{f,g\}\}_{\lambda}(u,v)=\frac 1{\lambda}\sin\left(\lambda 
\left(\partial_u\partial_{v'}-\partial_{u'}\partial_v\right)\right)
f(u,v) g (u',v')|_{u=u',v=v'}
\end{equation}
Here, $\lambda$ corresponds to the Planck constant and the 
Moyal bracket gives a structure of infinite dimensional algebra
on the space of functions on the torus generated by
\begin{equation}
e_{r,s}(u,v) = \frac 1{2\pi} e^{\imath (r u+ s v)}
\end{equation}
where $u,v\in [0, 2 \pi]$ and $r,s \in {\Bbb Z}$. 
This algebra is the trigonometric algebra of  Fairlie Fletcher
and Zachos\cite{Fairlie:1989qd}:
\begin{equation}
\{\{e_{r,s},e_{r',s'}\}\}_{\lambda}(u,v)
 = \frac 1{2\pi\lambda} \sin\left(\lambda
\left(r s'-r' s\right)\right) e_{r+r',s+s'}(u,v)
\end{equation}
which  also icludes the case $\lambda = \frac{2\pi}N$. This case
gives the $SU(N)$ algebra in the base ${\cal J}_{r,s}$:
\bea
[{\cal J}_{r,s},{\cal J}_{r',s'}]= 
-2\imath \sin\left(\frac{2\pi}{N} (r s'-r's) \right) {\cal J}_{r+r',s+s'}
\eea
if the $e_{r,s}$ functions are identified with $e_{r+k N, s+ m N}$
for $k,m\in {\Bbb Z}$.   
The Heisenberg group matrices ${\cal J}_{r,s}$ have been introduced 
by Weyl.

When $\lambda \ra 0$ (or $N\ra \infty$), 
we recover the Poisson algebra of the area preserving
transformations of the torus~\cite{Floratos:1989mh} 
\bea
\{e_{r,s},e_{r',s'}\}(u,v) =  ( r's -r s') \frac 1{2\pi}
    e_{r+r',s+s'}.
\eea

The Matrix model has various large $N$--limits. Up to now
it is not known how to get the quantum mechanics of supermembrane
starting from this model, even though, various compactifications indicate
that it has  membrane states as excitations.
We believe that the appropriate limit is the 't Hooft topological
expansion   of the $SU(N)$  YM--mechanics.
To this end, we shall determine what happens to the Heisenberg group
matrices ${\cal J}_{r,s}$ in this limit. We observe that these
matrices contain powers of the root of unity along two diagonals
so we start 
with  $\omega = e^{2\pi\imath \frac MN}$, ($M,N$
co-prime integers).
The correct large $N$--limit for $SU(N)$ is the inductive
one, i.e., $SU(N)\ra SU(N+1)\ra SU(N+2) \dots $ which we get
if we let $M,N\ra \infty$ with $M/N =$ constant. 
Note that the constant  $\hbar=2 \pi\frac MN$ can be identified
with  the flux of the
3-index antisymmetric gauge field per unit membrane area.  
The Weyl relations
become the  Heisenberg group relations for an infinite 
phase-space lattice and in the Fourier transform space
of both canonical variables the Matrix model describes 
a toroidal continuous membrane with Matrix commutators 
replaced by Moyal brackets~\cite{Kavalov:1996ab}.
Since the limit $\hbar\ra 0$ replaces the Moyal bracket 
by Poisson, we get from Moyal YM theory the membrane.
Higher order corrections to $\hbar$ can be represented 
as membranes with attached handles on the initial 
membrane which is determined by the $SU(N)$ chosen basis,  
in our case the torus.

 In this limit,
the light-cone gauge equations of motion
for the membrane 
\beq \ddot{X}_i = \{X_k,\{X_k,X_i\}\}; \;\; i,k =
1,\ldots,
 d-1\label{eom}
 \eeq
and the corresponding Gauss law $\{X_i,\dot{X}_i\} = 0$
are replaced by 
\bea \ddot{X}_i &= &\{\{X_k,\{\{X_k,X_i\}\}\;\}\}
\label{eom0}\\
\{\{X_i,\dot{X}_i\}\} &=& 0,\;\; i,k =
1,\ldots d-1.\label{GL}
 \eea
 
When the space of functions on the toroidal membrane is
replaced by the algebra of $N\times N$
matrices, the coordinates of the membrane become the
matrices $A_{i}(t)$, the velocity is the $SU(N)$ electric field
$E_i(t)= \dot A_i(t)$, and the magnetic field in three or
seven dimensions is $B_i(t) =\frac 12 f_{ijk}[A_j,A_k]$ where 
$f_{ijk}$ is the $\epsilon_{ijk}$ totally antisymmetric 
symbol in three dimensions and $\Psi_{ijk}$ the 
octonionic multiplication table in seven 
dimensions~\cite{Dundarer:1983fe}.

The Moyal bracket generalizes both  Poisson brackets and
 matrix commutators, so that one is tempted to consider  
 a system where 
the Poisson bracket is replaced by the Moyal one~\cite{Curtright:1997st}.
The question of the appearance of Moyal bracket for physical
reasons in the dynamics of membrane is up to now open. 
 We know that there are other limits of the
Matrix model, one  leads to perturbative  string field theory
\cite{Dijkgraaf:1998iz,Bonelli:1999pq}, and the Poisson limit in which the $SU(N)$
symmetry becomes the area-preserving diffeomorphism group. 
We believe that the physical origin of the Moyal bracket 
is due to the presence of the antisymmetric 
background field $C_{ijk}$ in the light-cone gauge  
which gives a `magnetic' flux (Hall effect), 
trasforming the surface of the membrane into a non-commutative
phase--space\cite{Chapline:1998ns}. This is true for open membranes
where the topological term of the action receives a contribution
from the boundary.

\section{Topological charge, Bogomol'nyi bound and Integrability.}

In order to explain the appearance of non-abelian electric-magnetic type
 of duality in the membrane theory, we recall that for YM--potentials
independent of space coordinates the self-duality equation in the gauge
$A_0=0$ is
\bea
\dot{A}_i&=&  \frac 12\epsilon_{ijk}[A_j,A_k],\; i,j,k =
1,2,3 \label{sd3YM}
\eea
According to ref\cite{Corrigan:1983th} the only non-trivial higher
dimensional YM self-duality equations exist in 8 space-time
dimensions which, for the 7-space coordinate independent potentials,
can be written (in the $A_0=0$ gauge) as
\bea
\dot{A}_i&=& \frac 12\Psi_{ijk}[A_j,A_k],\; i,j,k = 1,\cdots 7
\label{sd7} \eea
 where $\Psi_{ijk}$ is the multiplication table of
the seven imaginary octonionic units. 

It is now tempting to take the large $N$-limit  and replace the
commutators by Poisson (Moyal) brackets to obtain the
self-duality equations for membranes (non-commutative instantons
for the Moyal case).
In this limit we replace the gauge potentials $A_i$ by the membrane
coordinates $X_i$. Then, the 3-d system is\cite{Floratos:1989hf,Ward:1990nz},
 \bea \dot{X}_i&=&
 \frac 12\epsilon_{ijk}\{X_j,X_k\},\; i,j,k = 1,2,3 \label{sd3},
\eea 
while in seven space dimensions~\cite{Curtright:1997st,Floratos:1997ky}
 \bea
  \dot{X}_i&=& 
\frac 12\Psi_{ijk}\{X_j,X_k\},\; i,j,k = 1,\cdots, 7 \label{sd70}
 \eea
and correspondingly for the case of Moyal 
brackets in three dimensions\cite{Castro:1997hr,Garcia-Compean:1996bj}
and in seven dimensions\cite{Fairlie:1997vj}.
It is easy to see that the self-duality membrane equations, imply
the second order Euclidean-time, equations of motion in the
light-cone gauge as well as the Gauss law.

One striking feature of the self-duality membrane
equations is their simple geometrical
meaning\cite{Floratos:1989hf,Floratos:1997ky}.
 These equations state that the normal vector
at a point of the membrane surface and the velocity at the same
point are parallel (self-dual) or anti--parallel (anti-self-dual).
The possibility to write down self-duality equations  based on the
existence of vector cross-product comes from the existence of the
quaternionic and octonionic algebras.
Since these are the only existing division algebras the 3 and 7
dimensions are unique\cite{Dundarer:1983fe}\footnote{For other
approach to self-duality see also\cite{Ivanova:1995rc}.}. 
The validity of this geometrical statement could be
extended in a  general curved space-time background as a
definition of the self-dual membranes.

If one takes the limit where the commutator of matrices is
replaced by commutator of operators or the Moyal bracket,
then the self-duality equations become the Moyal Nahm or 
Moyal-Bogomol'nyi equations of \cite{Curtright:1997st}.

The membrane instantons carry a topological charge density
\cite{Biran:1987ae} which satisfies a Bogomol'nyi bound
\cite{Zachos:1997cc}:
 \bea \Omega(X) =
\frac{1}{3!}\epsilon^{abc}f_{ijk}X_a^iX_b^jX_c^k
\eea
 where
$X_a^i=\partial_{\xi_a} X^i$, $a,b,c = 1,2,3$ and
$f_{ijk}=\epsilon_{ijk}$ when $i,j,k = 1,2,3$ and
$f_{ijk}=\Psi_{ijk}$ for $i,j,k =1,\cdots, 7$. This topological
charge density defines the topological charge of the membrane \bea
Q&=&\frac 1{V_3}\int d^3\xi \Omega(X) \eea where $V_3$ is the
volume of the integration region. The topological charge $Q$ is an
integer and represents the degree of the map from the
 membrane  to its world volume. 
 We display
below the convenient representation of the topological charge
which will help us demonstrate that it is a lower bound of the
membrane action for topologically non-trivial membranes 
\bea
\Omega(X)=  \frac 12\dot{X}_if_{ijk}\{X^j,X^k\}
    =\frac 12  \{X^j,X^k\}^2
\eea
 where the self-duality equations as well as the properties of
$f_{ijk}$ in three and seven dimensions have been used. The topological
charge of the membrane can now be written as
\bea Q&=&
\frac 1{2 V_3}\int_M d^3\xi \{X^j,X^k\}^2
\eea
 The minimum
value of $Q$ ($Q=1$) is obtained for the membrane instanton
compactified on a 
world--volume torus,  $X_1=\sqrt{2}\sigma_1$, $X_2=\sqrt{2}\sigma_2$ and
$X_3= 2 t$.

The Euclidean action can be written as 
 \bea  S &=& \frac
1{ V_3}\int d^3\xi
 \left(\frac 12 {{\dot{X}_i^{2}}} + \frac 14\{X_j,X_k\}^2
\right) \eea {}From the inequality
$  (\dot{X}_i \pm 
 \frac 12 f_{ijk} \{X_j,X_k\})^2\ge 0$
we derive that,
\bea
 S \ge Q
 \eea
 and the equality holds only for the self-dual or
anti-self-dual membranes. So the self-dual or anti-self-dual
membranes are BPS Euclidean-time membrane world-volume solitons.
As we have seen in ref\cite{Floratos:1998zq}, the 3$-d$ and 7-d self-dual
solutions preserve 8 and 1 supersymmetries respectively or $1/2$
and $1/16^{th}$ of the supersymmetry  of the light-cone
supermembrane Hamiltonian. This is a direct consequence of the
above Bogomol'nyi bound and the $SO(3)$ and $G_2$ rotational space
symmetry of the above cases. 

The role of the membrane instantons
is important in developing a perturbative expansion.  
Configurations of the membrane around instantons cannot 
collapse to points or strings, because they have different topological
charge. The 3-index antisymmetric gauge field which is so crucial
for the uniqueness of the supermembrane Lagrangian participates
in the bosonic part through the Cern-Simons term. If  its  
vacuum expectation value is  non-zro and proportional to $\Psi_{ijk}$
(in the corresponding 7 dimensions), then the topological charge 
defined above, separates the functional integral into membrane
topological sectors.
Going now to the case of Moyal-Nahm equations, there is a
corresponding topological charge without an obvious  geometrical
meaning and the Bogomol'nyi bound is valid  in  
 this case too.
This bound is important for the stability of the corresponding
Moyal-Nahm instantons. 
Recent discussions on the role of
instantons  in non-commutative  YM theories  (non-commutative
instantons) imply that they can be considered as regularizations
of small size instantons in standard YM theories
(see e.g. \cite{Nekrasov:1998ss}).
The case of Moyal-Nahm equations could be considered as non-commutative
membrane instantons which regularize the Poisson or membrane case.

We now make few remarks on  the integrability of the self-dual
equations. 
The 3-d self-duality system has a Lax pair and an infinite number of
conservation laws
\cite{Floratos:1989hf,Ward:1990nz}.
In order to see this, we first rewrite the self-duality equations in the form
\begin{equation}
\dot{X}_+=i\{X_3,X_+\}, \;\;
\dot{X}_-=i\{X_3,X_-\}, \;\;
\dot{X}_3=\frac{1}{2}i\{X_+,X_-\}, \label{17c}
\end{equation}
where
\begin{equation}
X_{\pm}=X_1\pm iX_2 \label{18}
\end{equation}
The Lax pair  equations can be written as
\beq
\dot{\psi}=L_{X_3+\lambda X_-}\psi, \quad \dot{\psi}=L_{\frac{1}{\lambda}
X_{+}-X_3}\psi,\label{21a,b}
\eeq
where the differential operators $L_f$ are defined as
\beq
L_f\equiv i\left(\frac{\partial f}{\partial\phi}
\frac{\partial}{\partial\cos\theta}-\frac{\partial f}
{\partial\cos\theta}\frac{\partial}{\partial\phi}\right).\label{22}
\eeq
The compatibility condition of $(\ref{21a,b})$ is
\beq
[\partial_t-L_{X_3+\lambda X_-},\partial_t-L_{\frac{1}{\lambda}
X_{+}-X_3}]=0,\label{23}
\eeq
from which, comparing the two sides for the coefficients of the powers
$\frac{1}{\lambda},\lambda^0,\lambda^1$ of the spectral parameter $\lambda$, we
find $(\ref{17c})$. From the linear
system  $(\ref{21a,b})$, using the inverse--scattering method, one could
in principle construct all solutions of the self-duality equations.

The infinite number of conservation laws are derived as follows:
from the Cartesian formulation
\bea
\frac{d X_i}{d t}&=& \frac 12 \epsilon_{ijk}\{X_j,X_k\}
\eea
contracting with a complex 3-vector $u_i$ such that
\bea
u_i &=& \epsilon_{ijk}u_jv_k,
\eea
where $u_iu_i = 0$, and $v$ is another
 complex vector with $v_iv_i = -1$ and $u_iv_i=0$,
we find,
\bea
\frac{d u\cdot X}{d t}&=& \{u\cdot X, v\cdot X\}
\label{3dsd}
\eea
The latter is a Lax pair type equation,  which implies
\bea
\frac{d }{d t}\int d^2 \xi (u\cdot X)^n = 0
\eea

Applying the same method in seven dimensions with
two complex 7-vectors  $u_i, v_i$
such that $u_iu_i = 0$,  $v_iv_i = -1$ and $u_iv_i=0$, leads to the equation
\bea
\frac{d u \cdot X}{d t} &=& \{ u\cdot X, v\cdot X\} +
\frac 12\phi_{jklm} u_j v_k
                \{ X_l,X_m\}\label{prev}
\eea

The curvature tensor $\phi_{jklm}$ is defined as the dual of $\Psi_{ijk}$ in
seven dimensions. When equation (\ref{prev}) is restricted to three dimensions
we recover (\ref{3dsd}). We observe that the presence of the curvature tensor
is an obstacle for the integrability.
At this point, we may look for an
 extended definition of integrability
replacing the zero-curvature condition with the
 octonionic curvature one.
  We can restrict  the above equation
in particular subspaces of solutions where integrability appears.
One possibility is the factorization of the time~\cite{Floratos:1997ky}.

We conclude with a few remarks. In this note we have given arguments
that the Matrix model  describes 
a non-commutative YM theory for the supermembrane in the presence 
of background three-index  antisymmetric gauge fields. 
If this conjecture is true, it implies
that the excitations of this model in various compactifications are
also physical excitations  of the supermembrane.
So the supermembrane should contain 11-d supergravity at least in
weak coupling limits given by small radii of the compactification
manifolds. It is tempting to calculate correlation functions of 
membrane observables using the Matrix model and then take the 
large $N$-limit as was discussed in section 3.
 On the other hand, perturbation theory for the 
supermembrane could be defined through the  expansion in the parameter
$\hbar/N$, with $M/N\ra \hbar$ for $M,N\ra \infty$. In this expansion
all the topologies of the membrane appear as splitting and joining interactions
 The other known
large $N$--limit~\cite{Dijkgraaf:1998iz,Bonelli:1999pq}
  gives the string perturbation theory 
as a QM  sector of the supermembrane. 

As this work was written, we have been kindly informed that the Moyal
limit of the Matrix model has been studied in connection with the
higher derivative corrections to the Born-Infeld Lagrangians for
the D2--brane\cite{Cornalba:1999hn}. For a very recent, interesting
paper on D--branes in group manifolds, see\cite{Garcia-Compean:1999uw}

\vspace*{1cm}
{ One of us (EGF) would like to thank prof. Albert Schwarz for a valuable
discussion.}


\begin{thebibliography}{99}
\bibitem{Witten:1996im}
E.~Witten,
Nucl. Phys. {\bf B460} (1996) 335
hep-th/9510135.
\bibitem{Townsend:1996af}
P.K.~Townsend,
Phys. Lett. {\bf B373} (1996) 68
hep-th/9512062.
%
\bibitem{Bergshoeff:1987cm}
E.~Bergshoeff, E.~Sezgin and P.K.~Townsend,
Phys. Lett. {\bf B189} (1987) 75.
\bibitem{deWit:1999rh}
B.~de Wit,
hep-th/9902051.
\bibitem{Banks:1997vh}
T.~Banks, W.~Fischler, S.H.~Shenker and L.~Susskind,
Phys. Rev. {\bf D55} (1997) 5112
hep-th/9610043.
\bibitem{Ishibashi:1996xs}
N.~Ishibashi, H.~Kawai, Y.~Kitazawa and A.~Tsuchiya,
Nucl. Phys. {\bf B498} (1997) 467
hep-th/9612115.
\bibitem{Polychronakos:1997fw}
A.P.~Polychronakos,
Phys. Lett. {\bf B403} (1997) 239
hep-th/9703073.
\bibitem{Goldstone}
G. Goldstone, MIT-1980, Unpublished. 
\bibitem{Hoppe} J. Hoppe, MIT-Ph. D. Thesis, (1982)
{\it Elem. Part. Research Journal} Kyoto, {\bf 83} (1989/90) 3.
\bibitem{Matinian:1981dj}
S.G.~Matinian, G.K.~Savvidy and N.G.~Ter-Arutunian Savvidy,
Sov. Phys. JETP {\bf 53} (1981) 421.
\bibitem{Simon:1983jy}
B.~Simon,
Annals Phys. {\bf 146} (N.y.) 209.
\bibitem{Bergshoeff:1990hw}
E.~Bergshoeff, E.~Sezgin, Y.~Tanii and P.K.~Townsend,
Ann. Phys. {\bf 199} (1990) 340.
\bibitem{deWit:1988ig}
B.~de Wit, J.~Hoppe and H.~Nicolai,
Nucl. Phys. {\bf B305} (1988) 545.
\bibitem{deWit:1989ct}
B.~de Wit, M.~Luscher and H.~Nicolai,
Nucl. Phys. {\bf B320} (1989) 135.
\bibitem{Floratos:1989au}
E.G.~Floratos,
Phys. Lett. {\bf B228} (1989) 335.
\bibitem{FInew}  E.G. Floratos and J. Iliopoulos, to appear.
\bibitem{Connes:1998cr}
A.~Connes, M.R.~Douglas and A.~Schwarz,
JHEP {\bf 02} (1998) 003
hep-th/9711162.
\bibitem{Hofman:1998iv}
C.~Hofman and E.~Verlinde,
Nucl. Phys. {\bf B547} (1999) 157
hep-th/9810219.
\bibitem{Douglas:1999ge}
M.R.~Douglas,
hep-th/9901146.
\bibitem{Douglas:1998fm}
M.R.~Douglas and C.~Hull,
JHEP {\bf 02} (1998) 008
hep-th/9711165.
\bibitem{Ho:1998xh}
P.~Ho and Y.~Wu,
Phys. Rev. {\bf D58} (1998) 066003
hep-th/9801147.
\bibitem{Moyal}
J.E.~Moyal,
Proc. Cambridge Phil. Soc. {\bf 45} (1949) 99.
\bibitem{Zachos:1999wn}
C.~Zachos and T.~Curtright,
hep-th/9903254.
\bibitem{Fairlie:1997vj}
D.B.~Fairlie,
Mod. Phys. Lett. {\bf A13} (1998) 263
hep-th/9707190.
%
\bibitem{Schomerus:1999ug}
V.~Schomerus,
JHEP {\bf 06} (1999) 030
hep-th/9903205.
\bibitem{string99}
E. Witten, N. Seiberg, Talks in Strings 99, Potsdam July 1999, \\
{\tt http://string99.aei-potsdam.mpg.de}
\bibitem{cartier} P. Cartier, ``Quantum Mechanical Commutation Relations and
Theta Functions'' in {\sl Proc. Symp. Pure Mathematics}, {\bf 9}:
{\sc Algebraic--Discontinuous Groups}, AMS, Providence, RI (1966).
\bibitem{bi} 
 D. Mumford, {\sl Tata Lectures on Theta}, { I--III},
Birkh\"auser, New York (1986).
\bibitem{Floratos:1997wg}
E.G.~Floratos and G.K.~Leontaris,
Phys. Lett. {\bf B412} (1997) 35
hep-th/9706156.
%
\bibitem{Fairlie:1989qd}
D.B.~Fairlie, P.~Fletcher and C.K.~Zachos,
Phys. Lett. {\bf B218} (1989) 203.
%
\bibitem{Floratos:1989mh}
E.G.~Floratos, J.~Iliopoulos and G.~Tiktopoulos,
Phys. Lett. {\bf B217} (1989) 285.
%
\bibitem{Kavalov:1996ab}
S.R.~Das, A.~Dhar, G.~Mandal and S.R.~Wadia,
Mod. Phys. Lett. {\bf A7} (1992) 71
hep-th/9111021.
\\
A.~Kavalov and B.~Sakita,
Annals Phys. {\bf 255} (1997) 1
hep-th/9603024.
%
\bibitem{Dundarer:1983fe}
R.~Dundarer, F.~Gursey and C.~Tze,
J. Math. Phys. {\bf 25} (1984) 1496.
\bibitem{Curtright:1997st}
T.~Curtright, D.~Fairlie and C.~Zachos,
Phys. Lett. {\bf B405} (1997) 37
hep-th/9704037.
\bibitem{Dijkgraaf:1998iz}
R.~Dijkgraaf, H.~Verlinde and E.~Verlinde,
Nucl. Phys. Proc. Suppl. {\bf 68} (1998) 28.
%
\bibitem{Bonelli:1999pq}
G.~Bonelli, L.~Bonora, F.~Nesti and A.~Tomasiello,
hep-th/9905092.
\bibitem{Chapline:1998ns}
G.~Chapline and A.~Granik,
hep-th/9808033.
\bibitem{Corrigan:1983th}
E.~Corrigan, C.~Devchand, D.B.~Fairlie and J.~Nuyts,
Nucl. Phys. {\bf B214} (1983) 452.
\bibitem{Floratos:1989hf}
E.G.~Floratos and G.K.~Leontaris,
Phys. Lett. {\bf B223} (1989) 153.
\bibitem{Ward:1990nz}
R.S.~Ward,
Phys. Lett. {\bf B234} (1990) 81.
\bibitem{Floratos:1997ky}
E.G.~Floratos and G.K.~Leontaris,
Nucl. Phys. {\bf B512} (1998) 445
hep-th/9710064.
\bibitem{Castro:1997hr}
C.~Castro and J.~Plebanski,
J. Math. Phys. {\bf 40} (1999) 3738
hep-th/9710041.
\bibitem{Garcia-Compean:1996bj}
H.~Garcia-Compean and J.F.~Plebanski,
Phys. Lett. {\bf A234} (1997) 5
hep-th/9612221.
%
\bibitem{Ivanova:1995rc}
T.A.~Ivanova and A.D.~Popov,
Theor. Math. Phys. {\bf 102} (1995) 280.
%
\bibitem{Biran:1987ae}
B.~Biran, E.G.~Floratos and G.K.~Savvidy,
Phys. Lett. {\bf 198B} (1987) 329.
%
\bibitem{Zachos:1997cc}
C.~Zachos, D.~Fairlie and T.~Curtright,
hep-th/9709042.
%
\bibitem{Floratos:1998zq}
E.G.~Floratos and G.K.~Leontaris,
Phys. Lett. {\bf B428} (1998) 75
hep-th/9802018.
%
\bibitem{Nekrasov:1998ss}
N.~Nekrasov and A.~Schwarz,
Commun. Math. Phys. {\bf 198} (1998) 689
hep-th/9802068.
%
\bibitem{Cornalba:1999hn}
L.~Cornalba and R.~Schiappa,
hep-th/9907211.
\bibitem{Garcia-Compean:1999uw}
H.~Garcia-Compean and J.F.~Plebanski,
hep-th/9907183.
\end{thebibliography}
\end{document}